\documentclass[Letter]{aa} 
\usepackage{graphicx}
\usepackage{txfonts}
\begin{document}

   \title{Does the virial mass drive the intra-cluster light?}

   \subtitle{The relationship between the ICL and M$_{vir}$ from VEGAS}

   \author{R.Ragusa
          \inst{1,}\inst{2}, E.Iodice \inst{1}, M.Spavone \inst{1},
          M.Montes \inst{3,4},
           Duncan A.Forbes \inst{5},
           S.Brough \inst{6}, M.Mirabile\inst{7,8}, M.Cantiello \inst{7},
          M.Paolillo \inst{1,}\inst{2}
          \and P.Schipani\inst{1}
          }

   \institute{INAF Osservatorio Astronomico di Capodimonte, Salita Moiariello 16, 80131 Napoli, Italy
              \email{: rossella.ragusa@inaf.it}
         \and Università di Napoli “Federico II”, Via Cinthia 21, Napoli 80126, Italy.
             \email{: rossella.ragusa@unina.it}
             \and Instituto de Astrofìsica de Canarias, c/ Vìa Làctea s/n, E-38205 - La Laguna, Tenerife, Spain
             \and Departamento de Astrofìsica, Universidad de La Laguna, E-38205 - La Laguna, Tenerife, Spain  
            \and  Centre for Astrophysics and Supercomputing, Swinburne University, John Street, Hawthorn VIC 3122, Australia 
            \and School of Physics, University of New South Wales, NSW 2052, Australia
            \and INAF Osservatorio Astronomico di Teramo, via Maggini, I-64100, Teramo, Italy
            \and Gran Sasso Science Institute (GSSI), I-67100 L'Aquila, Italy}

   \date{Received XX,XX,XXXX; accepted  XX,XX,XXXX}


  \abstract
   {In this Letter we revisit the relationship between the fraction of the intra-cluster light (ICL) and both the virial mass and the fraction of Early Type Galaxies in the host halo. This is based on a statistically significant and homogeneous sample of 22 groups and clusters of galaxies in the local Universe ($z \leq 0.05$), obtained with the VST Early-type GAlaxy Survey (VEGAS).
   Taking advantage of the long integration time and large area of the VEGAS images, we are able to map the galaxy outskirts and ICL down to $\mu_g$ $\geq$ 29-30 mag/arcsec$^2$ and out to hundreds of kpc. With this data-set, we have expanded the sample of ICL measurements, doubling the previous measures available from the literature for z $\leq$ 0.05. 
   The main result of this work is the lack of any significant trend between the fraction of ICL and the virial mass of the host environment,  covering a wide range of virial masses ( $\sim$ $10^{12.5} \leq M_{vir} \leq 10^{15.5} M_{\odot}$), in agreement with some theoretical studies.
   Since the new data points are all derived with the same methodology and from the same observational setup, and all have comparable depth, the large observed scatter indicates an intrinsic variation in the ICL fraction.
   On the other hand, there is a weak relation between the fraction of ICL and the fraction of Early Type Galaxies in the host halo, where a larger fraction of ICL is found in groups and clusters of galaxies dominated by earlier morphological types, indicating a connection between the ICL and the dynamical state of the host system.}
   
  
   
   

\authorrunning{R. Ragusa et al.}
\titlerunning{The relationship between the ICL and M$_{VIR}$ from the VEGAS survey}
\maketitle

%

\section{Introduction}\label{sec:intro}

Since proposed and then observed by \citet{zwicky1937ApJ....86..217Z,zwicky1951PASP...63...61Z,Zwicky1952PASP...64..242Z,Zwicky1957PASP...69..518Z} in the Coma cluster, we know that there is an additional component in the light distribution of galaxy clusters: 
the intra-cluster light (ICL). 
The ICL is made of baryons (stars and globular clusters) gravitationally unbounded to any specific galaxy in the cluster, which is observed as a diffuse and very faint emission \citep[$\mu_V$ > 26.5 mag/arcsec$^2$,][]{Mihos2005}, 
extendeing out to several hundred of kpc from the cluster centre \citep[see][and references therein]{montes2019intracluster}.
According to the $\Lambda$CDM paradigm, the ICL is the product of the gravitational interactions (accretion and merging) involved in the formation of the most massive galaxies at the centre of groups and clusters, named as the brightest group (BGG) or cluster (BCG) galaxy. 
Therefore, the ICL is usually more concentrated around the BCG (or BGG), as confirmed by observations \citep[e.g.][]{Mihos2005,Arnaboldi2012}. 
%
Several formation channels have been proposed for the build-up of the ICL, which may all happen in the same environment: {\it i)} tidal stripping of satellites in the potential well of the BCG \citep{Rudick2009,Contini2014,Contini2019}, {\it ii)} disruption of dwarf galaxy members \citep{Purcell2007},
{\it iii)} major mergers with the BCG or BGG \citep{Murante2007,Conroy2007ApJ...668..826C},
{\it iv)} pre-processing in groups \citep{Mihos2005,Rudick2006} or {\it v)} from unbound stars formed in-situ \citep{Puchwein2010}.
%
The low surface brightness (LSB) features resulting from these processes (e.g. shells, tidal tails, stellar streams) also contribute, a few percent, to the total amount of ICL \citep[e.g.][]{gonzalez2013ApJ...778...14G,Presotto2014,contini2018MNRAS.479..932C}. 
Hence, the ICL represents the fossil record of the mass assembly in galaxies. As a consequence, 
the physical properties of the ICL (e.g. total luminosity, colors, stellar populations) 
allow us to constrain the formation channels for this component, 
the look back time and the dynamical state of the system \citep[see][and references therein]{Contini2021Galax...9...60C,montes2022NatAs...6..308M}.
%
Simulations agree that the bulk of the ICL forms in the redshift range $0\leq z \leq 1$, 
with the total amount of ICL increasing up by $\sim$ 20$\%$ to $z = 0$. In this context, one of the key parameters is the fraction of ICL, $f_{ICL}$ (Sec.\ref{subsec:ICL}).
%
%
%
From a theoretical perspective, more evolved clusters are expected to have larger values of $f_{ICL}$, since galaxies in the dense environment of galaxy clusters are expected to have experienced more gravitational interactions \citep[e.g.][]{Murante2007,Rudick2011,Martel2012,Contini2014}.
A way to constrain the physical processes that form the ICL is to see how $f_{ICL}$ correlates with the virial mass (M$_{vir}$)
of the host environment. Theoretical predictions report contradictory results. According to several works \citep{Sommer-Larsen2006,Monaco_2006,Henriques_2010, Rudick2011,Contini2014}, in the halo mass range M$_{vir} \simeq 10^{13}-10^{15}$~M$_\odot$, $f_{ICL}$ spans from 20\% to 40\% without any trend with M$_{vir}$. 
Conversely, increasing values of $f_{ICL}$, from 20\% up to 40\%, with incresing M$_{vir}$ are predicted in several simulations \citep{Lin_2004,Murante2007,Purcell2007,Pillepich2018,henden2019baryon,Ahad2022arXiv221014249A}.
An opposite trend is suggested by \citet{Cui_2013}, where a decreasing $f_{ICL}$, from $\sim$ 50\% 
to  $\sim$ 40\%, with increasing M$_{vir}$, is predicted.\\
%
Due to its low surface brightness nature, the detection and study of the ICL is a challenging task, which requires very deep
images, covering large areas around the center of the clusters or groups.
So far, the few available measurements of $f_{ICL}$ have prevented any conclusion on the correlation with 
M$_{vir}$ \citep[see][as reviews, and references therein]{Contini2021Galax...9...60C, montes2022NatAs...6..308M}.
In the local universe ($z\leq 0.05$), $f_{ICL}$ spans a wide range of values, from 10\% up to $\sim 50\%$ for any M$_{vir}$, therefore it seems that the two quantities are not correlated \citep{RAGUSA2021, Ragusa2022FrASS...952810R,montes2022NatAs...6..308M}. A similar result is obtained for ICL measurements at higher redshift \citep[$0.2 \leq z \leq 0.35$, e.g.][]{Sampaio-Santos2021}.
On the other hand, it seems that a mild trend is observed between $f_{ICL}$ and the fraction of the early type galaxies ($f_{ETGs}$=ETGs/[ETGs+LTGs]), i.e.
the fraction of ICL is larger in those environments dominated by early type galaxies \citep{DaRocha2008,RAGUSA2021,Ragusa2022FrASS...952810R}.
Even considering the few data points available, \citet{Iodice_2020} found that the $f_{ICL}$ seems
to increase with decreasing amount of neutral hydrogen (HI) in the host system, another tracer of the evolutionary state of the system \citep[e.g.][]{Kilborn2009}. However, these results need to be confirmed with larger data-sets. In particular, since the ICL content can be estimated using several methods \citep[see the review by][]{montes2022NatAs...6..308M} and, in addition, it depends on the depth of the data and on the adopted data analysis, 
the main requirements
needed to address the dependencies of the $f_{ICL}$ with M$_{vir}$ and the evolutionary state of the environments are a homogeneous and statistically significant sample of groups and clusters of galaxies, spanning the whole halo mass range covered by the theoretical predictions.\\
In this Letter we revisit the relationships between $f_{ICL}$ versus M$_{vir}$ and $f_{ICL}$
versus $f_{ETGs}$ based on a sample of 22 groups and clusters, covering
$10^{12.5} \leq M_{vir} \leq 10^{14.5} M_{\odot}$ in the local universe (i.e. $z\leq 0.05$), from the VST Early-type GAlaxy Survey \citep[VEGAS,][]{Capaccioli2015, IODICE2021Msngr.183...25I}.
The images are all acquired with the same observational setup, have comparable depths (within $\sim$ 0.5 mag) and are all analysed using the same methodology. This allows us to overcome many 
of the limitations mentioned above and provide more robust results on the connection between the ICL fraction and the properties of the host environment.
The Letter is organized as  follows: in Sec.~\ref{sec:obs} we present the VEGAS observations and the data reduction strategies used for this work. In Sec.~\ref{sec:method} we describe the data analysis and performed. Finally, in Sec.~\ref{sec:discuss} we illustrate our results and compare them with theoretical predictions.
We assume $H_0 = 73 $km s$^{-1}$ Mpc$^{-1}$, $\Omega_M = 0.3$, and $\Omega_{\Lambda}= 0.7$.

\section{Deep images from VEGAS: Observations $\&$ Data reduction}\label{sec:obs}

\begin{figure*}
    \centering
    \includegraphics[width=18cm]{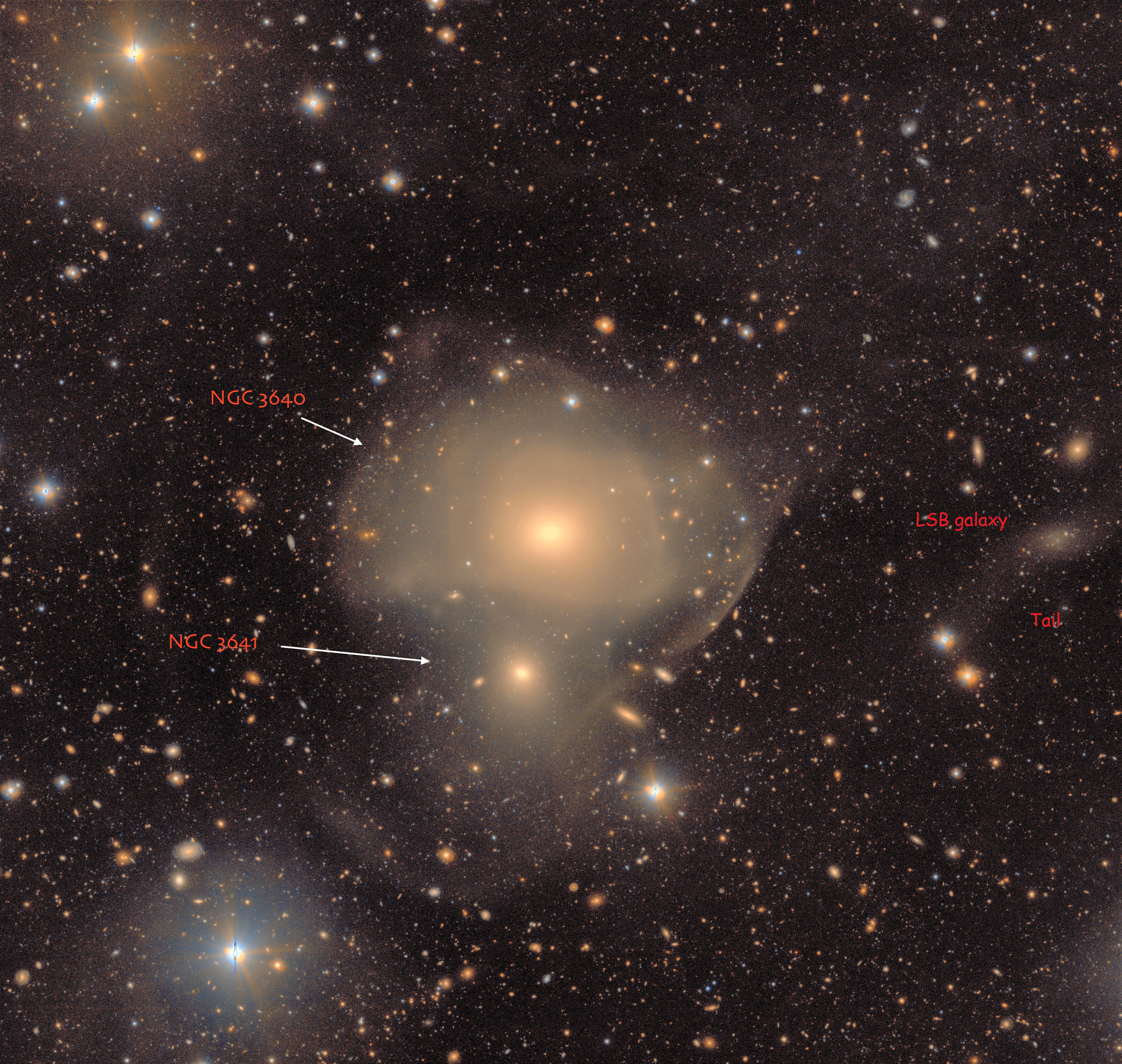}
     \caption{Color composite $(gri)$ VST image of the NGC~3640 group. The image size is $\sim$ 20 x 20 arcmin. North is up and East is to the left. The brightest group members are labelled in red on the image.
     The image is chosen as a target example for the VEGAS sample, given how clearly LSB features ( $\mu_g$ > 29 mag/arcsec$^2$), such as shells and tidal tails, are evident in the outskirts of the central galaxy. Moreover, on the West side of NGC~3640, a LBS galaxy is also detected, with a very faint tail towards the BGG (Mirabile et al., in prep.). For this target, the fraction of IGL plus diffuse stellar envelope, estimated in Sec.~ \ref{sec:method}, is $\sim$ 8\%, and comes mainly in the form of shells and tidal tails around the BGG.}
     \label{fig:composite}
       \end{figure*}

The VEGAS sample presented in this Letter is made of 22 targets in total, five of which have already been subject of previous works, covering the core of groups and clusters of galaxies in the lower redshift regime ($z\leq 0.05$). The new 17 targets reported here are listed in Tab.~\ref{tab:sample&igl}.
VEGAS is a multi-band, deep imaging survey \citep{IODICE2021Msngr.183...25I},
based on the observations acquired
with the {\it ESO VLT Survey Telescope (VST)}, a 2.6 meter optical telescope located at 
Cerro Paranal, Chile \citep{Schipani2012}. The VST is equipped with the wide field camera, OmegaCAM, with a field of view of $1^{\circ} \times 1^{\circ}$ and a resolution of $0.21$~arcsec~pixel$^{-1}$.
The data presented in this work were acquired in different visitor or service mode runs, in dark time and clear conditions. 
%
The data reduction is performed using the dedicated AstroWISE \citep[for more details see][]{McFarland2013,Venhola_2017} or VST-Tube \citep[][]{Capaccioli2015} pipelines, both are able to provide equivalent results.
A detailed description of the observing setup and data reduction for the VEGAS images 
have been presented in several papers \citep[e.g][]{Capaccioli2015,Iodice2016,Iodice_2020,Spavone2017b,Cattapan2019,RAGUSA2021}, 
to which we refer for a comprehensive description and, in particular, for the adopted tools and methods for the background removal, which is fundamental in such LSB studies.
All targets in our sample are observed in the {\it g} and {\it r} filters, some of them also have images acquired in the {\it u} and/or {\it i} bands. In this Letter, we focused on the ICL estimates derived in the {\it g} band, which is the most efficient OmegaCAM  filter \citep{kuijken2011Msngr.146....8K}. All {\it g}-band images have a total integration time of $\sim2.5$~hrs. 
The surface brightness depth at 5$\sigma$ over an area of the average seeing of FWHM $\sim$ 1 arcsec is  $\mu_g \sim 29-30$~mag/arcsec$^2$,
which is about six magnitudes fainter than the surface brightness of the night sky in the $g$ band at ESO-Paranal \citep[$\sim$ 22.5 mag/arcsec$^2$,][]{Desai_2012}. 
In Fig. \ref{fig:composite} we show the resulting sky-subtracted color composite VST image obtained for the NGC~3640 group.
This image reveals the plethora of shells and tidal tails detected
around the BGG, NGC~3640, and shows how the data acquisition and reduction perform well to unveil  
the LSB features in the galaxy outskirts and intra-group space.

\section{Data analysis}\label{sec:method}

The data analysis used to estimate the ICL for all the targets in our sample is based on the well-tested surface photometry of the galaxies, which is described in detail in our recent paper \citep{RAGUSA2021}. 
The two main preliminary steps are the {\it i)} removal of the bright stars in the field, and {\it ii)} estimate of the limiting radius of the surface photometry. In short, to account for the scattered light from the bright stars in the field that might affect the estimate of the ICL, they are modelled and subtracted from the original image. During this step, the core of the group (or cluster) is also masked out to $\sim$ 10-20 R$_{eff}$ of the BCGs, to avoid accounting for part of the ICL in the scattered light. 
On the stars subtracted images, we estimated the limiting radius of the photometry ($R_{lim}$, hereafter). This corresponds to the outermost radius with respect to the center of the target, where the galaxy's light blends into the residual background fluctuations. 
This task is performed by fitting the light distribution in the frame in circular annuli, centred on the BCG (or BGG).

\subsection{Measuring the ICL}\label{subsec:ICL}  

 According to the two-phase formation scenario of galaxy formation, the inner and brighter part of the BCG is formed
fist, with a surface brightness profile well reproduced by a S{\'e}rsic law. During the accretion phase, the bounded stellar envelope and unbounded ICL are built around the BCG.
Given their similar accreted origin, based on photometry alone, it is not possible to unambiguously separate the ICL 
from the stellar envelope, since 
the transition between these two components occurs very smoothly and sometimes is completely indistinguishable, 
as both components show a similar surface brightness profile.
For this reason, the $f_{ICL}$ is commonly defined as the total contribution from the diffuse and unbounded intra-cluster baryons plus the stellar envelope around galaxies.
There are different photometric methods in the literature used to separate the ICL from the BCG \citep[see][as review]{montes2022NatAs...6..308M}.
One of the proposed and more reliable methods is based on the 1D or 2D multi-component 
decomposition of the BCG light distribution, where the brightest regions of the BCG can be 
distinguished from the stellar envelope and ICL
\citep[e.g][]{Gonzalez2007,Kravtsov2018AstL...44....8K,Zhang2019ApJ...874..165Z,kluge2020ApJS..247...43K,montes2021buildup}.
In detail, by fitting the light distribution of the BCG with empirical laws, the transition region between the gravitationally bound and brightest regions of the BCG and the stellar envelope plus the ICL is constrained. 
 As commented before, due to the difficulty separating between BCG, bounded stellar halo and unbounded ICL, the total amount of ICL reported here includes the contribution from the stellar envelope and the unbound diffuse light. Therefore, the total amount of ICL must be considered as an upper limit.

%

We have adopted this method to study the stellar halos in BCGs and to estimate the contribution of ICL from VEGAS images. 
It is  described and applied in several published papers \citep{Cattapan2019,Iodice_2020,Spavone_2020,Raj_2020,RAGUSA2021,Ragusa2022FrASS...952810R}. Below, we summarize the main steps that lead to the estimate of $f_{ICL}$, in the $g$ band. 

\begin{itemize}

\item{\it 1D multi-components decomposition of the BCGs}: we performed the 1D multi-component decomposition of the BCGs’ azimuthally-averaged surface brightness profiles, adopting the procedure introduced by \citet{Spavone2017b}. This provides the estimate of the transition radius ($R_{tr}$, hereafter) in each galaxy. The $R_{tr}$ is the distance from the galaxy centre where the contribution from the galaxy outskirts (i.e. stellar envelope plus diffuse light) starts to dominate the total light distribution. Hence, based on the isophote fit, we derived the 2D-model of the BCG's light distribution out to $R_{tr}$ and subtracted it from the parent image. 
%
%
  \item{\it Residual images}: 
  on the residual image obtained in the previous step, using the same multi-component fitting approach, we have also modelled the light from all the other galaxy members (out to $R_{lim}$),
  including those objects that do not show a prominent contribution of diffuse light in the outskirts. All the 2D models have been subtracted from the parent image.
  The resulting residual image 
  traces the spatial distribution of the stellar envelopes plus the ICL in the group or cluster, and this is used to compute 
  its total luminosity (L$_{ICL,g}$). 
  The value of L$_{ICL,g}$ for all targets in our sample is reported in Tab. \ref{tab:sample&igl}.
%
%
   \item{\it Estimate of the ICL fraction}: $f_{ICL}$ is defined as the ratios L$_{ICL,g}$ / L$_{tot,g}$, where L$_{ICL,g}$ is the total luminosity of the ICL and L$_{tot,g}$ is the 
   total luminosity of the group or cluster in the $g$  band, including the ICL. L$_{tot,g}$ is obtained by summing up the total luminosity derived for all cluster (or group) members plus
   L$_{ICL,g}$. All the above values for the targets in our sample are listed in Tab.~\ref{tab:sample&igl}. 
 The error estimate on L$_{ICL,g}$ and L$_{tot,g}$ takes into account the uncertainties on the photometric calibration, the RMS in the background fluctuations and the poissonian error assumed for the total integrated flux\footnote{The error on the total magnitude is computed as follow: $err_{mag}= \sqrt{\{2.5/[flux \times \ln(10)]\}^2 \times [(err_{flux}+err_{sky})^2] +  err_{zp}^2}$, where $err_{sky}$ is the RMS on the sky background, $err_{zp}$ is the error on the photometric calibration, and $err_{flux}=\sqrt{flux/N-1}$ is the poissonian error on the integrated flux from a total number of N pixels used in the isophote fit.}.
 
\end{itemize}
%


The surface photometry for all galaxies in all targets of our sample (i.e. surface brightness profiles, colors, integrated magnitudes and total luminosity), derived by the isophote fitting,  will be presented and discussed in a forthcoming paper (Ragusa et al., in preparation). In this Letter we focus on the main outcome of this analysis: the estimate of the ICL for such a large and homogeneous sample of groups and clusters of galaxies.

\section{Discussion}\label{sec:discuss}

The main goal of this Letter is to revisit the relationship between the ICL fraction with {\it i)} the halo mass of the environment 
(M$_{vir}$), shown in Fig.~\ref{fig:massaviriale}, and {\it ii)} with the fraction of the evolved galaxy members (the ETGs), shown 
in Fig.~\ref{fig:ETGLTG}. 
To date,
both of these relationships have remained uncertain, and theoretical predictions do not provide any constraint 
(see Sec.~\ref{sec:intro}). 
As far as ICL estimates are concerned in the halo mass range  $10^{12.5} \leq M_{vir} \leq 10^{15.5}$~$M_{\odot}$, with the new sample presented in this work, we significantly enlarged the statistics for the study of both relationships, doubling the previous literature estimates with a homogeneous sample 
(see Sec.~\ref{sec:method}). Therefore, all possible biases in the results due to observational setup and/or methodology are minimized.
%


\subsection{ICL versus M$_{vir}$ }\label{sec:Mvir}

\begin{figure*}
    \centering
    \includegraphics[width=16cm]{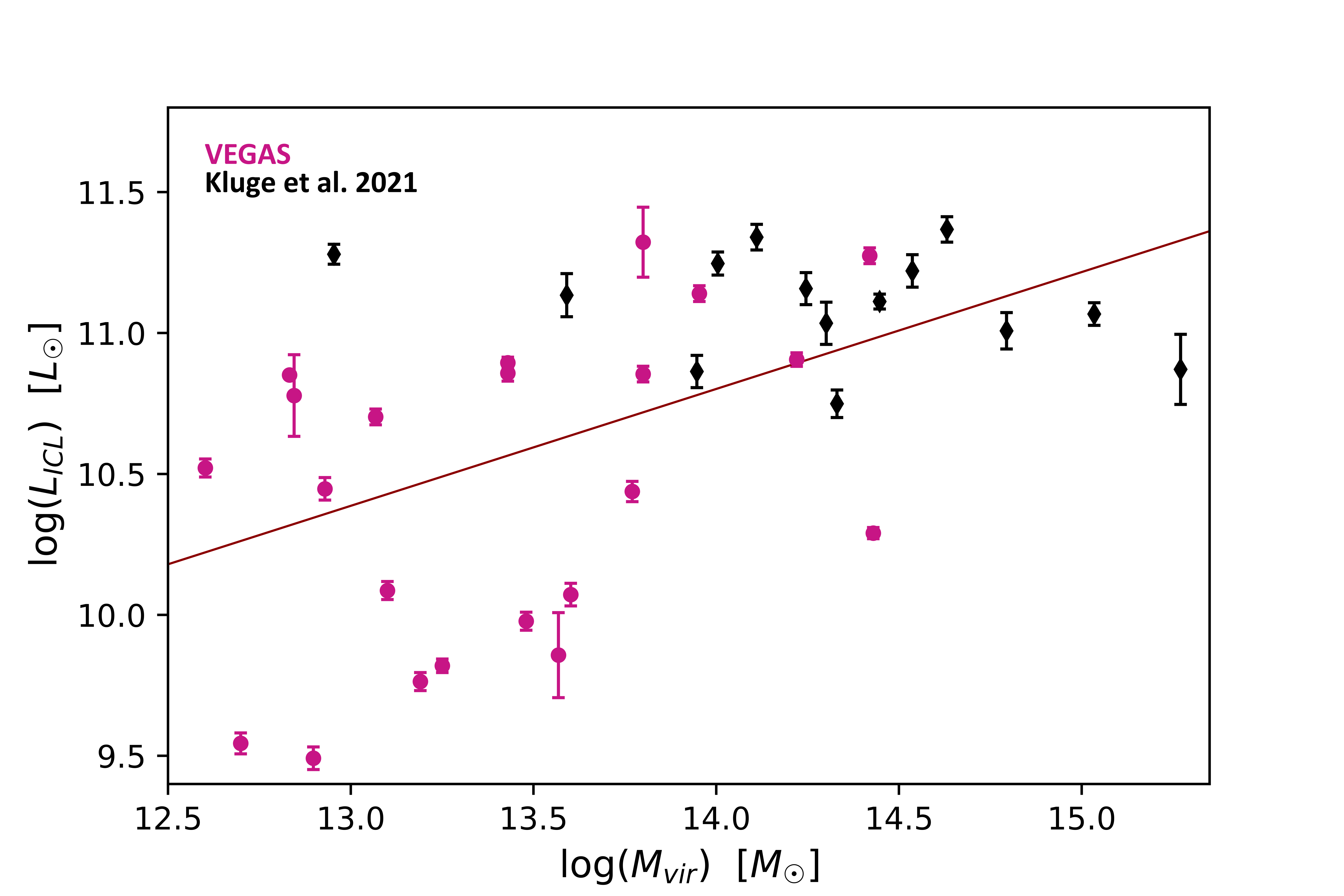}
    \includegraphics[width=16cm]{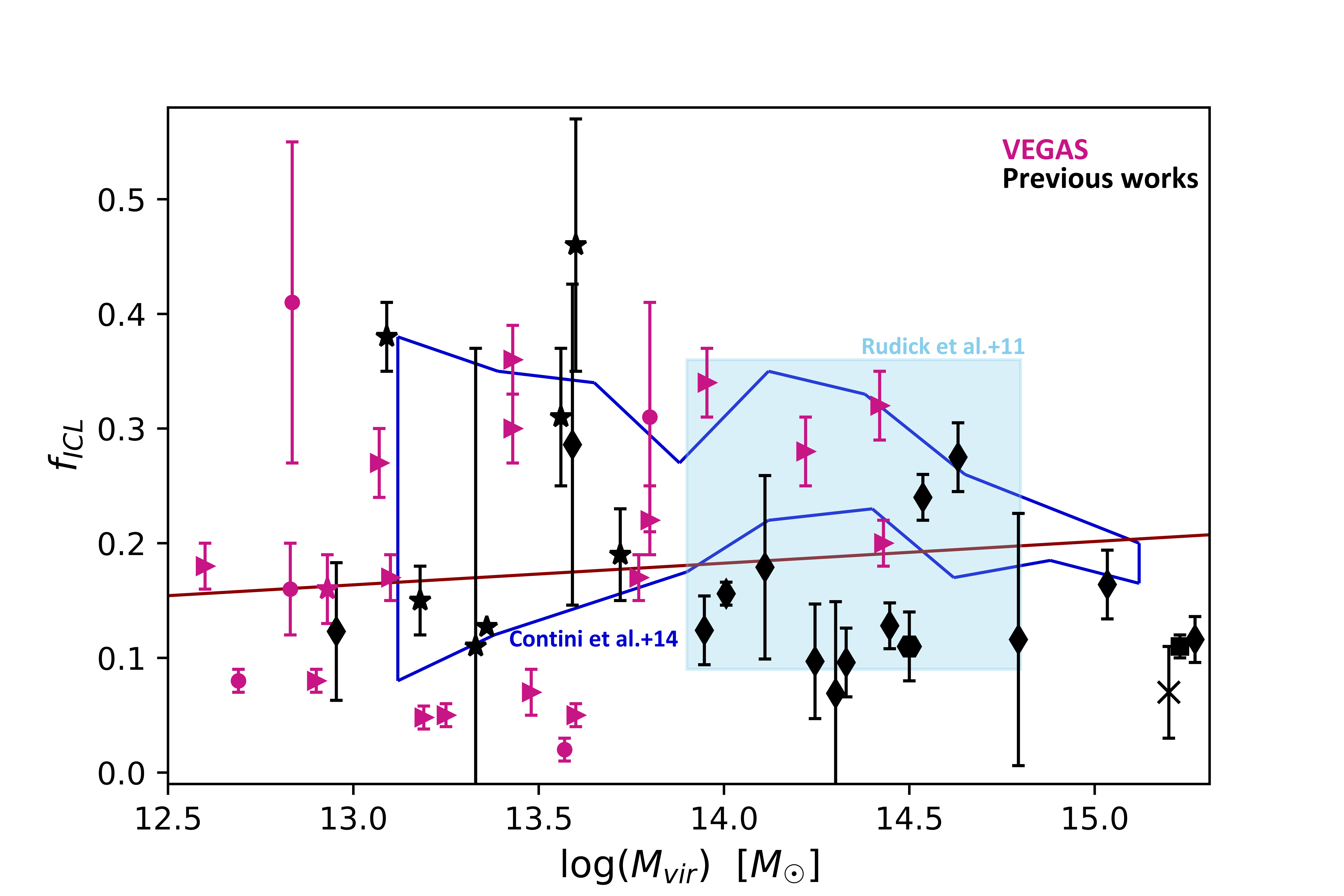}
     \caption{{\it Top panel:} Luminosity of the ICL component as a function of the M$_{vir}$. The best fit for the linear correlation is shown by the 
     dark-red solid line. The best fit equation is log($L_{ICL}$) = (0.41 $\pm$ 0.11)~*~log($M_{vir}$) - (5 $\pm$ 1.5), with R$^2$ = 28\% and p-value=0.
     {\it Lower panel:} ICL fraction ($f_{ICL}$) versus $M_{vir}$ obtained for the VEGAS targets (magenta symbols), compared with the values of $f_{ICL}$ available in the literature, for targets at 
     z $\leq $ 0.05 (black points). These are the compact groups from  \citet[][stars]{DaRocha2005,DaRocha2008,Pildis_1995,Poliakov2021,RAGUSA2021}, the Coma cluster \citep[][octagon]{Jim_nez_Teja_2019}, the Virgo cluster \citep[][cross]{Mihos2017} and the Abell~85 cluster \citep[][square]{montes2021buildup}. The clusters of galaxies from \citet{kluge2021ApJS..252...27K} are also included in this plot (diamonds). 
      The virial masses for the cluster in the latter sample are from \citet{kluge2020PhDT........30K}, when available. For those systems with no estimation of the viral mass, we used the velocity dispersion of the member galaxies reported in \citet{kluge2020PhDT........30K} to estimate it following \citet{Munari2013MNRAS.430.2638M}.
     The solid line indicates
     the best fit for the linear correlation. The best fit equation is $f_{ICL}$ = (0.02 $\pm$ 0.02)~*~log($M_{vir}$) - (0.07 $\pm$ 0.34), with R$^2$ = 1.6 \% and p-value= 0.47.
     The theoretical prediction obtained by \citet{Contini2014} (blue contours) and \citet{Rudick2011} (light blue area) are included for comparison.
     }
     \label{fig:massaviriale}
       \end{figure*}

\begin{figure*}
    \centering
    \includegraphics[width=18cm]{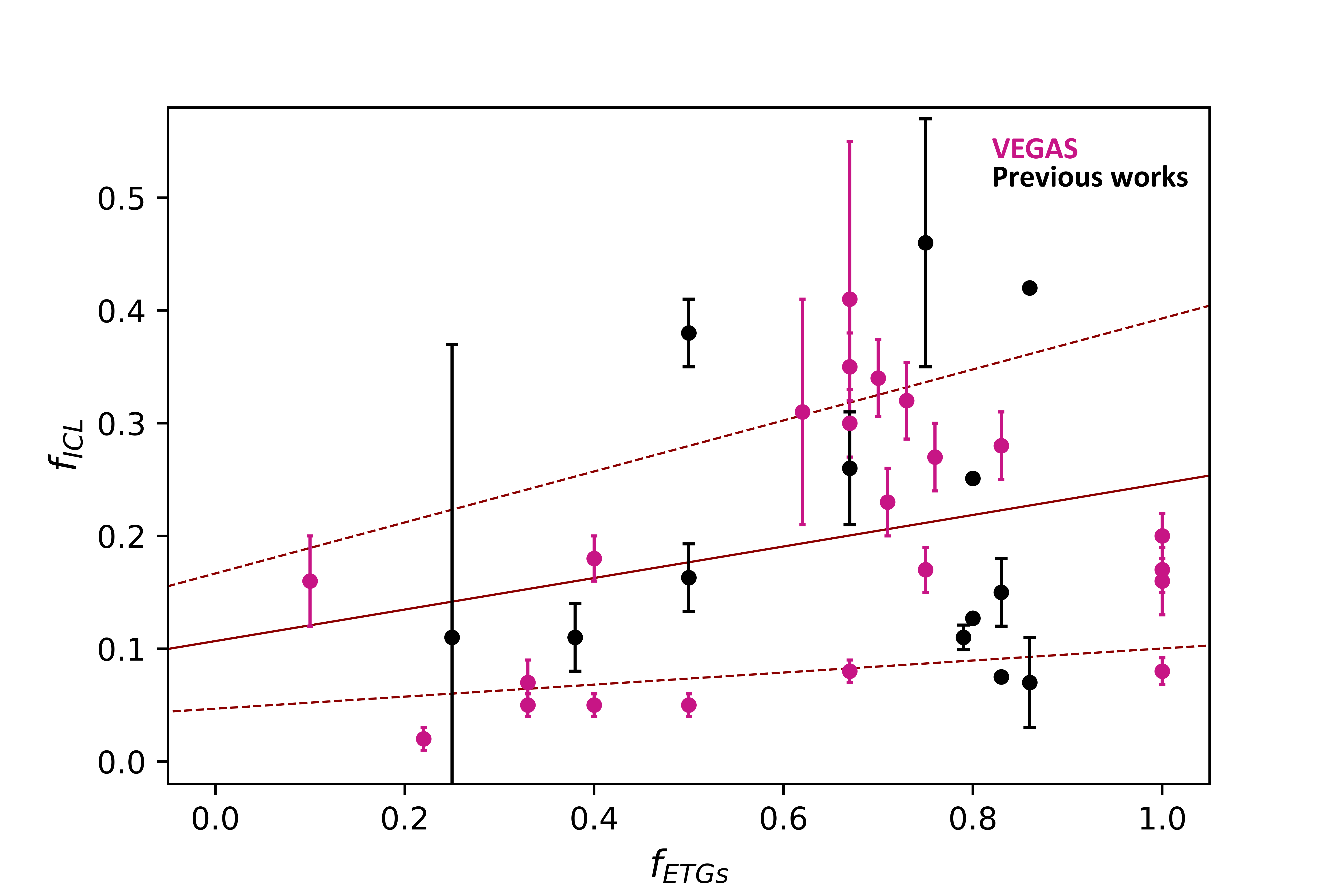}
     \caption{ICL fraction ($f_{ICL}$) as a function of the $f_{ETGs}$ . 
     The values for the ICL obtained from VEGAS are marked in magenta and the other estimates for groups and clusters of galaxies available in the literature in black (all the references as in Fig.~\ref{fig:massaviriale}). 
     A weak trend exists between the two quantities, as indicated by the dark-red solid line, which reflects the best fit for the linear correlation. The dashed red lines mark the 1$\sigma$ significance range of that correlation.
     The best fit equation is $f_{ICL}$ = (0.14$\pm$ 0.08)~*~$f_{ETGs}$  - (0.11 $\pm$ 0.06), with R$^2$ $\sim$ 10 \% and p-value = 0.12.
     }
     \label{fig:ETGLTG}
       \end{figure*}

In Fig.~\ref{fig:massaviriale} we plot $L_{ICL}$ (upper panel) and  $f_{ICL}$ (lower panel) versus M$_{vir}$ of the environment, 
for a sample of 46 systems at z $\leq$ 0.05. 
The VEGAS sample accounts for 22 targets. The remaining 24 targets 
come from the literature, where $f_{ICL}$  was obtained with similar methods we used in this work and in the optical wavelength bands, to avoid introducing any bias in the relationship. 
%
%
As expected, a mild trend is found for $L_{ICL}$ versus $M_{vir}$ (see Fig.~\ref{fig:massaviriale}, upper panel), as more satellite galaxies, lead to the formation of more ICL \citep[e.g.][]{Sampaio-Santos2021,kluge2021ApJS..252...27K}.
Conversely, the trend disappears for $f_{ICL}$ versus $M_{vir}$ (see Fig.~\ref{fig:massaviriale}, lower panel).
In particular, we found that a large $f_{ICL}$ ($\sim 30-40$\%) can be present in both loose and less massive groups 
of galaxies, like NGC~5018 (M$_{vir}$ < 10$^{13} M_{\odot}$), and in rich and massive clusters of galaxies,
such as the Antlia cluster (M$_{vir}$ $\sim$ 10$^{14} M_{\odot}$).
Similarly, a small $f_{ICL}$ (0-15$\%$) has been detected in both compact and less massive groups,
( 10$^{13} M_{\odot}$ < M$_{vir}$ <  10$^{13.5} M_{\odot}$), and in very rich and massive clusters, such as Abell~85 and Coma ( 10$^{15} M_{\odot}$ < M$_{vir}$ <  10$^{15.5} M_{\odot}$).
Based on this plot, in agreement with previous results by \citet{RAGUSA2021, Ragusa2022FrASS...952810R} and \citet{montes2022NatAs...6..308M}
on a smaller sample of data, a large scatter is observed for $f_{ICL}$ versus M$_{vir}$. The two quantities do not show any strong correlation, as confirmed by the linear regression best fit to the sample.
%
Since most of the data in Fig. \ref{fig:massaviriale} come from a data-set with the same depth and methodology, it is reasonable
to conclude that the observed scatter in this plot is intrinsic, and it
cannot be due to inconsistencies in the data acquisition and analysis.
As addressed in Sec.~\ref{sec:intro}, theoretical predictions provide different and sometimes contradictory
results on the relationship between $f_{ICL}$ and M$_{vir}$. The large range of estimates now available
from this work, allow us to provide stringent constraints on the most reliable scenario connecting the two
physical quantities.
In the lower panel of Fig.~\ref{fig:massaviriale}, we have superposed the theoretical predictions by \citet{Contini2014} and \citet{Rudick2011}.
Neither works predict a significant trend between the fraction of ICL and the halo mass of the host environment, in agreement with the new observations. In particular, in their mass ranges and resolutions, they found that $9\% \leq f_{ICL} \leq 40\%$, 
consistent with the measured range presented here.
The absence of a strong physical connection between the fraction
of ICL and the halo mass of the host environment found here 
would suggest that 
the formation mechanisms of the ICL do not depend on the potential 
well where it is located, and
they are equivalently efficient at the scale of both groups and clusters \citep[see also][]{Canas2020}. In particular, the bulk of the ICL formation
might happen in groups of galaxies, that subsequently join the cluster potential, and 
a smaller fraction of ICL is formed later in the cluster environment.
In support of this argument, the two sub-groups of the Antlia cluster, presented in this Letter, 
have comparable values of $f_{ICL}$ to that estimated for the whole cluster.
This would point toward a different efficiency of the several formation channels proposed for the build-up of the ICL, and in particular towards the pre-processing in groups as the main channel for the build-up of the bulk of the ICL.




\subsection{ICL fraction versus $f_{ETGs}$}\label{sec:ETG/LTG}

In Fig. \ref{fig:ETGLTG} we show $f_{ICL}$ as a function of the ratio 
between early-type and the total number of bright galaxies ($f_{ETGs}$, with $M_{B} \leq 16$) in 
the host environment, for the VEGAS sample and for those targets in the literature for which an estimate is available. 
As already pointed out by \citet{RAGUSA2021}, this larger sample hints that systems with a higher number of ETGs show a larger fraction of ICL. Fitting all the points
with a linear regression, we find a weak trend between $f_{ICL}$ and $f_{ETGs}$. However, within 1$\sigma$ from the best fit, the scatter is also quite large. The $f_{ICL}$ ranging from 5\% up to 40\% in all range of $f_{ETGs}$. Similarly, we found that groups or clusters of galaxies with a comparable fraction of ICL,
in the range 5-15\%, show very different $f_{ETGs}$, from 0 to 1. 
The weak trend with the $f_{ETGs}$ found here supports the idea that the fraction of ICL is connected with the efficiency of the
several formation channels, i.e. gravitational interactions, like tidal stripping and merging, which are more frequent in those environments that are more dynamically evolved, i.e that show a larger $f_{ETGs}$ ratio.





\section{Conclusions}

In this Letter we have presented new estimates of the intra-cluster light fraction, based on deep images from VEGAS, for a homogeneous and large sample (22 targets) of groups and clusters of galaxies, in the halo mass range
$10^{12.5} \leq M_{vir} \leq 10^{14.5}$~M$_{\odot}$, for z $\leq$ 0.05.
We have found that the $f_{ICL}$ ranges from $\sim 5\%$ up to $\sim 50\%$. Such values are consistent with the predicted $f_{ICL}$ from several theoretical studies.
We have investigated the relation between $f_{ICL}$ and the halo mass of the target and also with its fraction of early-type galaxies, by combining our targets with estimates for additional 24 targets, available from previously published works.
The large sample allows us to confirm that, although as expected in those systems with more satellite galaxies, the $L_{ICL}$ increases with  $M_{vir}$, the $f_{ICL}$ does not depend on the $M_{vir}$ of the host environment, in agreement with predictions from \citet{Contini2014} and \citet{Rudick2011}. 
This has fundamental implications for the assembly history of groups and clusters. The $f_{ICL}$ depends on the efficiency of the formation mechanisms of this diffuse light, being larger in those systems with a large number of more evolved galaxies, as also suggested from the mild relation between the $f_{ICL}$ and $f_{ETGs}$ found in this work. The lack of a strong correlation between $f_{ICL}$ and $M_{vir}$ leads us to conclude that the bulk of the ICL is assembled at the group scale and then incorporated into the clusters, during the groups infall.
It is worth noting that in the $f_{ICL}$ versus M$_{vir}$ plane, we cover well the low M$_{vir}$ range ($10^{12.5} \leq M_{vir} \leq 10^{13.5}$~$M_{\odot}$), which was so far poorly covered in previous works.
The lack of more ICL estimates for the most massive clusters of galaxies ($10^{14.5} \leq M_{vir} \leq 10^{15.5}$~$M_{\odot}$), for z $\leq$ 0.05, is mainly due to the lack of deep images covering the large area needed for such a systems, which require extended mosaics with the available wide-field telescopes for surveys.
The upcoming observing facilities, aimed at covering large portions of the sky down to the LSB regime \citep[such as the 10-year Legacy Survey of Space and Time, which will take place at the Vera C. Rubin Observatory, e.g.][]{Montes2019MNRAS.482.2838M,Brough2020arXiv200111067B}, will be able to provide also accurate ICL estimates in this halo mass range.


\begin{table*}
\setlength{\tabcolsep}{1.2pt}
\centering
\caption{New unpublished VEGAS sample of groups and clusters and ICL estimates.} 
\begin{tabular}{lcccccccccc}
\hline\hline
 Target & Environ. &R.A &    Decl.& D & m$_{ICL,g}$ &    L$_{ICL,g}$ &  R$_{vir}$ & M$_{vir}$& L$_{ICL,g}/L_{tot,g}$& $f_{ETGs}$  \\
     &&[J2000] & [J2000] & [Mpc]& [mag] & [$10^{10}L_{\odot}$] & [Mpc] & [$10^{13}M_{\odot}$]&&\\
     (1)&(2)&(3)&(4)&(5)&(6)&(7)&(8)&(9)&(10)&(11)\\
    \hline
Antlia & cluster &10:30:00.7 & -35:19:31.7 &40.0& 10.27 $\pm $ 0.07& 18.80 $\pm $ 1.20 & 1.28 & 26.3 & 0.32$\pm $ 0.03& 0.76 \\
HCG~86 & compact gr. &19:52:08.8& -30:49:32.7 & 82.0&13.90$\pm $ 0.10 & 2.80 $\pm $ 0.30&  0.41 & 0.85 & 0.16 $\pm $ 0.03& 1.00\\
NGC~596/584 &group & 01:32:12.1 & -06:56:54.6 &27.0& 13.21$\pm $ 0.08 & 0.58 $\pm $ 0.04& 0.50  & 1.55 & 0.05$\pm $ 0.01&0.40 \\
NGC~1453 &group & 03:46:27.3 & -03:58:07.6&55.5& 14.00 $\pm $ 0.10 & 1.20 $\pm $ 0.10& 0.86  &4.00  &0.05$\pm $ 0.01&0.50\\
NGC~1553 &group &  04:16:10.5 & -55:46:48.5& 17.3 &10.86 $\pm $ 0.09 & 2.70 $\pm $ 0.20 & 0.80  & 5.89 & 0.17$\pm $ 0.02&0.75\\
NGC~3100 & group &10:00:40.8 & -31:39:52.4 & 36.0& 13.70$\pm $ 0.06 &0.66$\pm $ 0.04 & 0.52  & 1.78 &0.05 $\pm $ 0.01&0.33\\
NGC~3258 &Antlia sub-gr. &10:28:53.6 & -35:36:19.9 & 41.0&11.70$\pm $ 0.07 & 5.00$\pm $ 0.30& 0.46  & 1.17& 0.27$\pm $ 0.03&0.76\\
NGC~3268 & Antlia sub-gr.  & 10:30:00.7 & -35:19:31.7&40.0 & 10.60 $\pm $ 0.07 & 13.80 $\pm $ 0.90& 0.90  & 8.99 & 0.34 $\pm $ 0.03&0.70\\
NGC~3379 & group &10:47:49.6 & +12:34:53.9 &10.2& 10.72 $\pm $ 0.08 & 1.22$\pm $ 0.09 & 0.47  & 1.26 & 0.17$\pm $ 0.02 & 1.00\\
NGC~3640 & group &11:21:06.9 &  +03:14:05.4 & 18.0 & 13.20$\pm $ 0.10 & 0.30 $\pm $ 0.03& 0.40  & 0.79 & 0.08$\pm $ 0.01 & 1.00\\
NGC~3923 & group &11:51:01.7 & -28:48:21.7 &23.0& 10.01 $\pm $ 0.05 & 7.80$\pm $ 0.40 & 0.60  & 2.69 & 0.35 $\pm $ 0.03&0.67\\
NGC~4365 & Virgo sub-gr. & 12:24:28.3 & +07:19:03.6 & 21.3& 10.79 $\pm $ 0.08  & 3.30$\pm $ 0.20 & 0.32  & 0.40 & 0.18$\pm $ 0.02&0.40\\
NGC~4636 & Virgo sub-gr.  &12:42:49.9 & +02:41:16.0 & 20.0&  12.01 $\pm $ 0.08& 0.95$\pm $ 0.07& 0.63  & 3.02 &0.07$\pm $ 0.02&0.33\\
NGC~4697 & Virgo sub-gr.& 12:48:35.9 & -05:48:02.7 & 18.0& 11.00 $\pm $ 0.05& 1.95$\pm $ 0.09 & 1.29  & 26.9 & 0.20 $\pm $ 0.02& 1.00 \\
NGC~5044 & group &13:15:24.0 & -16:23:07.9&32.0 & 10.84 $\pm $ 0.07& 7.20 $\pm $ 0.50 &  0.80 & 6.31 &0.22$\pm $ 0.03& 0.71 \\
NGC~5846 & group &15:06:29.3 & +01:36.20.3  &25.0& 10.17 $\pm $ 0.06 & 8.10$\pm $ 0.40& 1.10  & 16.6 &0.28$\pm $ 0.03&0.83\\
NGC~6868 & group &20:09:54.1 & -48:22:46.4 &41.0 & 11.36 $\pm $ 0.07 & 7.20$\pm $ 0.50  & 0.60  & 2.69 & 0.30$\pm $ 0.03& 0.67\\

\hline
\end{tabular}
\tablefoot{Col 1 reports the name of the BGG-BCG and in Col 2 the environment in which the target lies. In Cols 3 and 4 are listed the celestial coordinates of the group-cluster center. Col 5 reports the distance of the BCG. In Cols 6 and 7 are given respectively the magnitude and luminosity of the ICL component, in the $g$ band, from our work. In Cols 8 and 9 are listed the virial radius and virial mass inside the virial radius, respectively. In Col 10 is the fraction of the  ICL component with respect to the total luminosity of the group-cluster in the $g$ band, and in Col 11 the fraction of the ETGs with respect to the total number of galaxies, taking into account the brighter galaxies studied in each system. All the magnitudes listed in the table are corrected for Galactic extinction using the extinction coefficients provided by \citet{Schlafly_2011}. To compute the L$_{\odot,g}$ we use the prescriptions from \citet{blanton2003ApJ...592..819B}, with M$_{\odot,g}$ = 5.45 mag. For the M$_{vir}$ (R$_{vir}$) we assumed the following values:  NGC~3923, NGC~4636, NGC~5044 are from \citet{Brough2006}, NGC ~6868, NGC~596/584, NGC~3640, NGC~4697, NGC~5846 are from \citet{Gourgoulhon1992A&A...255...69G}, NGC~4365 is from \citet{samurovic2014A&A...570A.132S}, HCG~86 is from \citet{Coziol_2004}, NGC~3379 is from \citet{Karachentsev2015AstBu..70....1K}, NGC~3258, NGC~3268 and Antlia cluster are from \citet{calderon2020MNRAS.497.1791C}, NGC~3100 is from \citep{Tempel2016A&A...588A..14T}, and NGC~1453 is from \citep{Pandya2017ApJ...837...40P}. To 
estimate all the distances of this work we used the heliocentric
radial velocity, given by NED (NASA IPAC Extragalactic Database), and H$_{0}$ = 73 km s$^{-1}$ Mpc$^{-1}$ \citep{Riess2018}. }
    \label{tab:sample&igl}
\end{table*}


\begin{acknowledgements}
This study benefits from the observations gathered at the European Southern Observatory (ESO) La Silla Paranal Observatory within the VST Guaranteed Time Observations, Programme IDs: 090.B-0414(B),090.B-0414(D),091.B-0614(A),091.B-0614(B),092.B-0623(B),094.B- 0496(A),094.B-0496(B), 094.B-0496(D), 095.B-0779(A),096.B-0582(B),097.B-0806(A), 0101.A-0166(B),103.A-0181(A),0104.A-0072(B).
Authors acknowledge financial support from the VST project (P.I. P. Schipani) and from the INAF-OACN. Authors are very grateful to M. Kluge for providing the ICL estimates from the data set he studied. MM acknowledges the Project PCI2021-122072-2B, financed by MICIN/AEI/10.13039/501100011033, and the European Union “NextGenerationEU”/RTRP. 
 Authors wish also to thank the anonymous referee for all suggestions, which helped to improve this manuscript.
\end{acknowledgements}

\bibliographystyle{aa}
\bibliography{IGL}
\end{document}